\documentclass[preprint,12pt]  {elsarticle}
\usepackage{amssymb}
\usepackage{epsfig}
\usepackage{ifthen}
\usepackage{amsmath}
\usepackage{multicol}
\usepackage{bbm}
\usepackage[boxed,algosection,linesnumbered]{algorithm2e}
\usepackage{color}

\usepackage{epsfig}
\usepackage{ifthen}
\usepackage{amsfonts}
\usepackage{amssymb}
\usepackage{amsmath}
\usepackage{amscd}
\usepackage{amsthm}
\usepackage{fancyvrb}
\usepackage{epsfig}
\usepackage{mathtools}
\usepackage{epsfig}
\usepackage{bbm}
\usepackage{fullpage}
\usepackage[boxed,algosection,linesnumbered]{algorithm2e}
\begin{document}
\begin{frontmatter}
\title{On Equivalent Color Transform and Four Coloring Theorem}
\author[]{Wenhong Tian$^{a}$}
%\author[1]{GuoZhong Li}
%\author[1]{Xinyang Wang}
%\author[1]{Qin Xiong}
%\author[1]{YaQiu Jiang}
% \author[3]{Adel Nadjaran Toosi}
% \author[3]{Rajkumar Buyya}
\address[1]{School of Information and Software Engineering, \\University of Electronic Science and Technology of China (UESTC)}
%\address[2]{BigData Research Center of UESTC}
%\address[3]{CLOUDS Lab, Dept. of Computing and Information Systems, The University of %Melbourne, Australia}
%\author{}
%\address{}
\begin{abstract}
In this paper, we apply an equivalent color  transform (ECT) for a minimal $k$-coloring of any graph $G$.  It  contracts each color class of the graph to a single vertex  and produces a complete graph $K_k$ for $G$ by removing redundant edges between any two vertices.  Based on ECT, a simple proof for  four color theorem for planar graph is then proposed.\\
\end{abstract}
\begin{keyword}
Four color theorem\sep Equivalent color transform (ECT)
\end{keyword}
\end{frontmatter}
% \linenumbers

\section{Introduction}
Four Coloring Theorem (FCT) states that  every planar map is four colorable so that no two adjacent regions have the same color, given  any separation of a plane into contiguous regions in the map.  Proving four colors theorem is a very hard and challenging task since the first statement of the four color theorem in 1852.\\
The four color theorem was proved in 1976 by Kenneth Appel and Wolfgang Haken \cite{Appel1977a} \cite{Appel1977b} with computer assistance. It was the first major theorem to be proved using a computer. 

%Initially, their proof was not accepted by all mathematicians because the computer-assisted proof was infeasible for a human to check by hand %(Swart 1980). Since then the proof has gained wider acceptance, although doubts remain (Wilson 2014, 216�222).

To dispel remaining doubt about the Appel-Haken proof, a simpler proof using the same ideas  still with computer-assistance was published in 1996 by Robertson, Sanders, Seymour, and Thomas \cite{Robertson1996b} and the part which required hand verification was not as tedious as in Appel-Haken proof.  A quadratic algorithm to 4-color planar graph was proposed in \cite{Appel1989} by Appel and Haken. An efficient algorithm for four-coloring planar graph was also proposed in \cite{Robertson1996a}.  In this paper, we apply an equivalent color  transform (ECT) for planar graph and propose a simpler proof for four color theorem.
%Additionally, in 2005, the theorem was proved by Georges Gonthier with general purpose theorem proving software.

\section{Methods}

%\textbf{Theorem 1. Four Coloring Theorem (FCT)}: states that,no more than four colors are required to color the regions of a planar map so that no two adjacent regions have the same color, given  any separation of a plane into contiguous regions in the map. \\

%\textbf{Fermat's Last Theorem (FLT) }    states that:  for all $n$ greater than 2, there do not exist  $x$, $y$, $z$ such that 

%\begin{equation}
%x^n+y^n=z^n
%\end{equation}

% where $x$, $y$, $z$, $n$ are all positive integers.\\

\textbf{Definition 1. Equivalent Color Class (ECC).} For coloring a simple graph, node $i$ and $j$ can have the same color if they are non-adjacent, node $i$ and $j$ are called in equivalent color class.\\

\textbf{Definition 2. Equivalent Color Graph (ECG)}  is a composite graph, where  composite nodes are formed by moving those non-adjacent nodes in original graph together and edges are kept the same as in the original graph.\\

\textbf{Definition 3. Equivalent Color Transform (ECT)}  is  contracting each equivalent color class in ECG to a single vertex and keep just one edge between any adjacent vertex. 
%This will produce a complete simple graph $K_k$ from ECG. \\

There may be more than one way to form ECG for any given graph. Figure 1 shows an example of 6-node simple graph (a) and its corresponding ECG (b) and ECT (c).

\begin{figure} [htp!]
\begin{center}
%\hfill
{\includegraphics [width=0.65\textwidth,angle=-0] {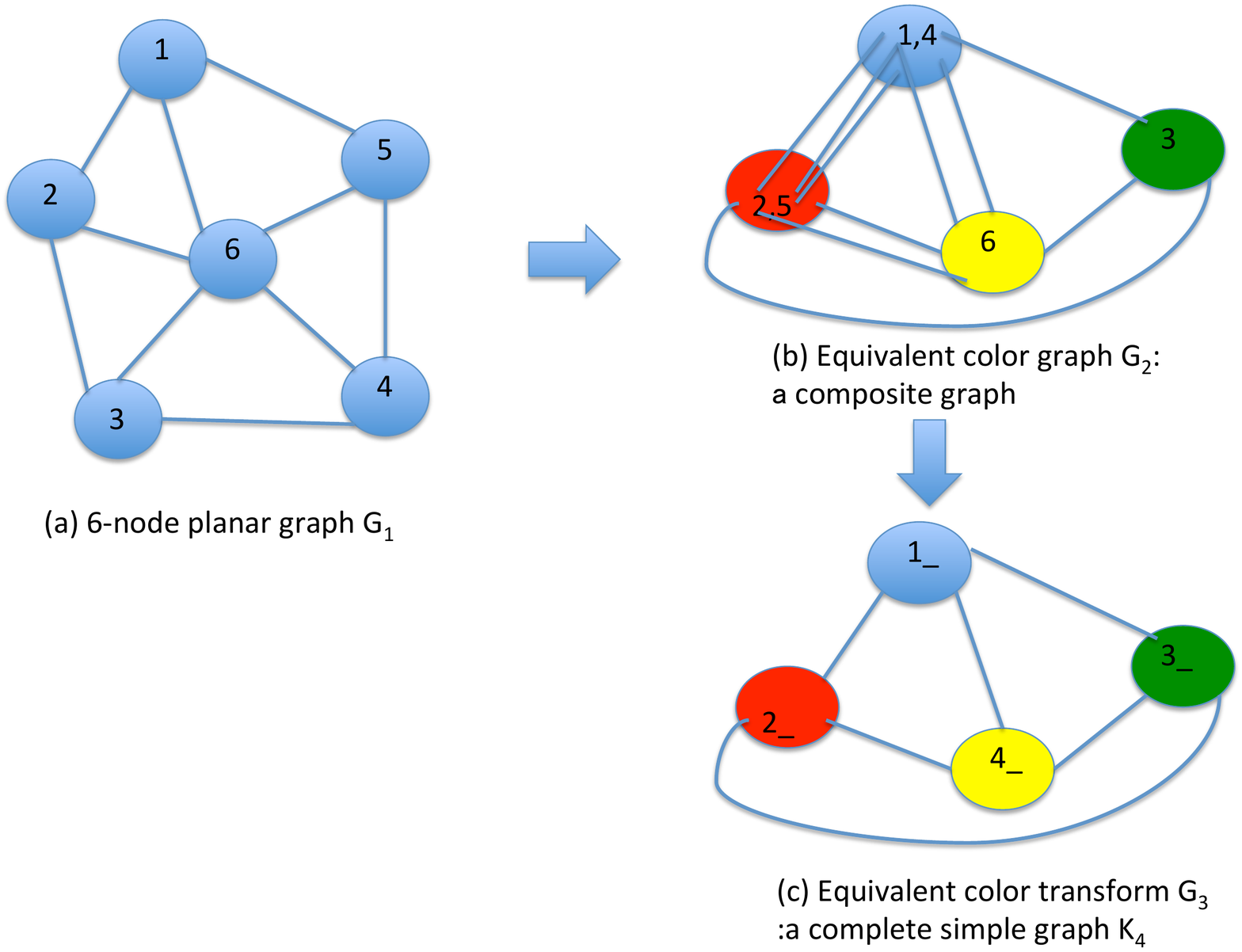}}
%\hspace*{\fill}
% \hfill \includegraphics [width=0.65\textwidth,angle=-0]{RunningTime.eps}\hspace*{\fill}
\caption{An instance of ECG and ECT}
\end{center}
\end{figure}

\textbf{Definition 4. Planar Graph}  is a graph that can be embedded in the plane, i.e., it can be drawn in such a way that no edges cross each other \cite{wikiplanar}.\\

\textbf{Lemma 1. There  exists an efficient ECT for planar graph.}  \\

There are exact algorithms to determine if a graph is 4-colorable such as one suggested in \cite{Formin2007}; there are also efficient algorithms such as one suggested in \cite{Robertson1996a} for 
efficiently four-coloring a planar graph. Notice that the RSST(abbreviated by the first letter of four authors in  \cite{Robertson1996a}) algorithm has computational complexity of $O(n^2)$, where $n$ is the number of nodes in the graph.  We therefore can conduct ECT efficiently by using RSST algorithm or similar algorithm.  \\

%Notice that exact algorithm to for 4 coloring the planar graph

\textbf{Theorem 1.  For a planar graph, there exist an ECT which does not change the planarity of the graph. } \\

\begin{proof}
Building ECG just moves the place of original nodes and does not alter any node or edge or their connection relationship, so we can apply RSST algorithm to find an ECT, which  do not change the planarity of the original graph. \\
%Based on ECG, ECT just reduce the number of nodes and some edges in original planar graph, therefore it does not change but keep  the planarity of the original graph.
\end{proof}
Notice that there may be more than one ECG and ECT for any given planar graph; we can apply RSST algorithm and similar algorithm to find an ECG and ECT, which do not change the planarity of the graph. To do this, we may need running RSST algorithm a few times until the ECG and ECT are found to keep the planarity of the original graph.

\textbf{Theorem 2.  A $k$-coloring graph which needs the minimum number of $k$ colors,  can be converted to a complete simple graph $K_k$ by ECT}.  \\

\begin{proof}
We prove by contradiction. Let us assume the original $k$-coloring graph $G_1$ is converted to a simple graph $G_2$ with $k$ nodes by ECT and $G_2$ is not complete graph. Since $G_2$ is not complete graph, there must exist at least two nodes $n_i$ and $n_j$ which are not adjacent. According to the definition of ECC and ECG, $n_i$ and $n_j$ can be contracted to a single vertex (by equivalent color class and ECT) and a new graph with ($k$-1) nodes is formed.  Therefore there only need ($k$-1) colors for $G_1$ and $G_2$, this contradicts the fact that $G_1$  is $k$-coloring and needs the minimum $k$ colors.  %Therefore $G_2$ becomes $K_{k-1}$
\end{proof}

\textbf{Lemma 1.  Wagner's Theorem. A finite graph is planar if and only if it does not have $K_5$ or $K_{3,3}$ as a minor \cite{wikiplanar}.}\\

\textbf{Theorem 3.  Planar graph  is four colorable}.  \\
\begin{proof}
We prove by contradiction. Set $k$ as the minimum number of colors needed for coloring planar graph $G$. Let assume the planar graph $G$ needs more than four colors, i.e., $k>4$.  From Theorem 2, we know that $G$ can be converted to a complete graph $K_k$ ($k>$4) in this case.  This means $G$ is transformed by ECT to a complete graph $K_k$ ($k>$4) which is not planar  because any complete graph with more than 4 nodes will have $K_5$ as a minor and  is not planar (by Lemma 1). But this contradicts the fact that $G$ is planar and there exits an ECT which does not change the planarity of the original graph (by Theorem 1). So $k\leq 4$ and planar graph is four colorable. 

\end{proof}%\section{Conclusions and Future Work}

\section*{Acknowledgments}
This research is sponsored by the National Natural Science Foundation of China (NSFC) (Grand Number:	61672136).  The manuscript is also posted online at https://arxiv.org/

\bibliographystyle{elsarticle-num}

\end{document}